\documentclass[10pt]{iopart}
 
\usepackage{hyperref}
\usepackage{graphicx}
\usepackage{iopams}
\usepackage{bbold}
\usepackage{xcolor}
\usepackage[xcolor,final]{changes}
\definechangesauthor[color=red]{GS}
\definechangesauthor[color=blue]{VS}

\newcommand{\ket}[1]{|#1\rangle}
\newcommand{\bra}[1]{\langle#1|}

\newcommand\deletedfootnote[1]{%
    \ifthenelse{\boolean{Changes@optiondraft}}
    {\footnote{\deleted{#1}}}
    {}}
\makeatother

\begin{document}
\title[Universal entanglement loss induced by angular uncertainty]{Universal entanglement loss induced by angular uncertainty}

\author{Giacomo Sorelli$^1$, Vyacheslav N. Shatokhin$^1$ and Andreas Buchleitner$^{1,2}$}
\address{$1$ Physikalisches Institut, Albert-Ludwigs-Universit\"at, 
Hermann-Herder-Stra\ss{}e 3, D-79104 Freiburg, Germany}
\address{$2$ Freiburg Institute for Advanced Studies, Albert-Ludwigs-Universit\"at, Albertstr. 19, 79104 Freiburg i.Br.}

%\noindent{\it Keywords\/}: entanglement, orbital angular momentum, uncertainty relation

\begin{abstract}
% single paragraph around 200 words
 \replaced{We study the diffraction of two maximally entangled qubits encoded into spatial modes of light with orbital angular momenta $\pm \hbar l_0$, on angular apertures.
Exploiting the uncertainty relation between angular position and momentum, we quantify the diffraction-induced entanglement loss, for different $l_0$, as a function of the angular uncertainty $\Delta \phi$ induced by the apertures.
We demonstrate that the degradation of entanglement follows a universal law parametrised by the product $l_0\Delta \phi$ alone.}
{We study the diffraction of two photonic qubits, which are initially maximally entangled in their orbital angular momenta, on angular apertures.  
Using the uncertainty relation between angular position and momentum, we show that the diffraction-induced entanglement losses exhibit universal behaviour.}
\end{abstract}

%\submitto{\JOPT}

%\ioptwocol
\section{Introduction}
\label{sec:intro}
Photons with a well defined orbital angular momentum (OAM) \cite{Allen:92}, also known as ``twisted photons'', are equipped with an infinite-dimensional degree of freedom which qualifies them as 
carriers for large amounts of (quantum) information \cite{Mair:2001,DiLorenzo:2010,Dada2011,Wang:2012,Eckert:1991,Gr_blacher_2006}, but also as versatile objects to test fundamental aspects of quantum mechanics in a well-defined, continuous degree of freedom with periodic boundary conditions \cite{Barnett:1990,Franke_Arnold_2004,JhaPRA2008, Jack_NJP_2008,Jack_JOPT_2011}.

\replaced{L}{Since l}arge capacities are associated with high resolution, here with respect to the OAM states' distinctive feature, their helical phase front structure\replaced{. T}{, t}his implies strong sensitivity of OAM-encoded information with respect to disturbances along the propagation direction.
Such perturbations may be of deterministic or random character\replaced{. A}{, with a}tmospheric turbulence \added{is} a prominent example \replaced{of random disturbance}{for the latter} \cite{Anguita:08,Sorelli_2019}, resulting in severe entanglement degradation \cite{Sorelli_2019,Ibrahim:2013,Roux:2015,Leonhard:2018,Leonhard2015,Bachmann:2019}. 

In our present contribution, we focus on \replaced{a deterministic}{the former} scenario, specifically on the fate of OAM entangled states when diffracting upon angular apertures.
Since the latter localize the photon(s) in a sharply defined interval of the angular degree of freedom, this entails a complementary uncertainty in the conjugate angular momentum which -- in turn -- defines the OAM states.
\added{This broadening of the OAM distribution can be interpreted quantum mechanically, by invoking the uncertainty relation between angular position and momentum \cite{Barnett:1990, JhaPRA2008, Jack_NJP_2008}, which was experimentally verified in \cite{Franke_Arnold_2004,Jack_JOPT_2011}.}
We \replaced{therefore expect}{can therefore anticipate that} OAM-entanglement evolution upon diffraction on an angular aperture \replaced{to}{must} be fundamentally controlled by \replaced{this}{the quantum mechanical} uncertainty relation\deleted{between angular position and momentum}:
initially well-defined, non-separable coherent superpositions of angular-momentum biphoton states experience a diffraction-induced spreading of the OAM basis upon scattering, which will generically reduce separability. 
We now set out to turn this into a quantitative statement.

The paper is organized as follows:  In section \ref{Sec:angular_apertures}, we discuss how to describe  diffraction on angular apertures numerically, followed by a recollection of basic facts on angular position-momentum uncertainty in section \ref{Sec:Angle-OAM-uncertainty}.
Section \ref{Sec:entanglement_losses} generalises these results for the diffraction of entangled biphoton state, and establishes a universal entanglement decay law under diffraction upon angular apertures.
Section \ref{Sec:conclusion} concludes the paper.

\section{Diffraction of twisted photons on angular apertures}
\label{Sec:angular_apertures}
We consider single photons with an OAM of $\hbar l_0$ and the spatial profile of a Laguerre-Gaussian (LG) mode with radial index $p=0$ \cite{andrews_babiker}:
\begin{equation}
u_{l_0} (\rho,\phi,z=0) = \mathcal{N} (\rho/w)^{|l_0|}e^{il_0\phi}e^{-\rho^2/w^2},
\label{LG}
\end{equation}
where $\rho$ and $\phi$ are, respectively, the radius and the azimuthal angle in the $x-y$ plane, while $w$ is the beam radius at the beam waist $z=0$, and $\mathcal{N}$ is a normalization constant, whose specific form is irrelevant here.

We seek the diffracted \replaced{field}{image} $\psi_{l_0}(\rho,\phi,z)$ \replaced{originating from}{of} the mode $u_{l_0}(\rho,\phi, 0)$  at $z>0$ behind an angular aperture located in the $z=0$ plane. 
The effect of such a diffracting screen is described by multiplication of the input mode \eref{LG} with the screen's transmission function $t(\rho,\phi)$ for modified Kirchhoff boundary conditions \cite{goodman}: $\psi_{l_0}(\rho,\phi,0) = u_{l_0}(\rho,\phi, 0)t(\rho,\phi)$. 
For our present problem, the transmission function is given by the product of a\deleted{n azimuthal} Gaussian \added{function of the azimuthal variable}, which ensures the angular confinement, and a radial super-Gaussian,
\begin{equation}
t(\rho,\phi)=\frac{(\lambda/\pi)^{1/4}}{\sqrt{\mathrm{erf}(\pi \sqrt{\lambda})}} e^{-\lambda\phi^2/2}e^{- (\rho/a)^{2m}},
\label{cake-slice}
\end{equation}
where $\lambda$ determines the angular width of the aperture, while $a$ is the radius of the super-Gaussian, which we specify with the power $m=12$ \footnote{This particular value was chosen small enough to smoothen the edges of the diffracting screen and, at the same time, large enough to limit the size of the edge region.}. 
The super-Gaussian removes edge effects encountered under standard Kirchhoff boundary conditions \cite{diff_paper, Fischer:07}, while the azimuthal Gaussian allows to match the wave function of intelligent states\deletedfootnote{That is, states for which the uncertainty relation \eref{uncertainty_relation} turns into an equality, as discussed in section \ref{Sec:Angle-OAM-uncertainty}.}
\added{namely states that saturates the uncertainty relation for angular position and angular momentum, as discussed in section \ref{Sec:Angle-OAM-uncertainty}}
\deleted{for the angular position-angular momentum uncertainty relation \eref{intelligent}}. \added{
We will not consider sharp edge apertures \cite{JhaPRA2008,Jack_NJP_2008} here, since these induce qualitatively identical results while being numerically harder to deal with (note, however, our discussion in section \ref{Sec:Angle-OAM-uncertainty}).}
%%%%%%%%%%%%%%%%%%%%%%%%%%%%%%%%%%%
\begin{figure*}
\centering
\includegraphics[width=0.7\textwidth]{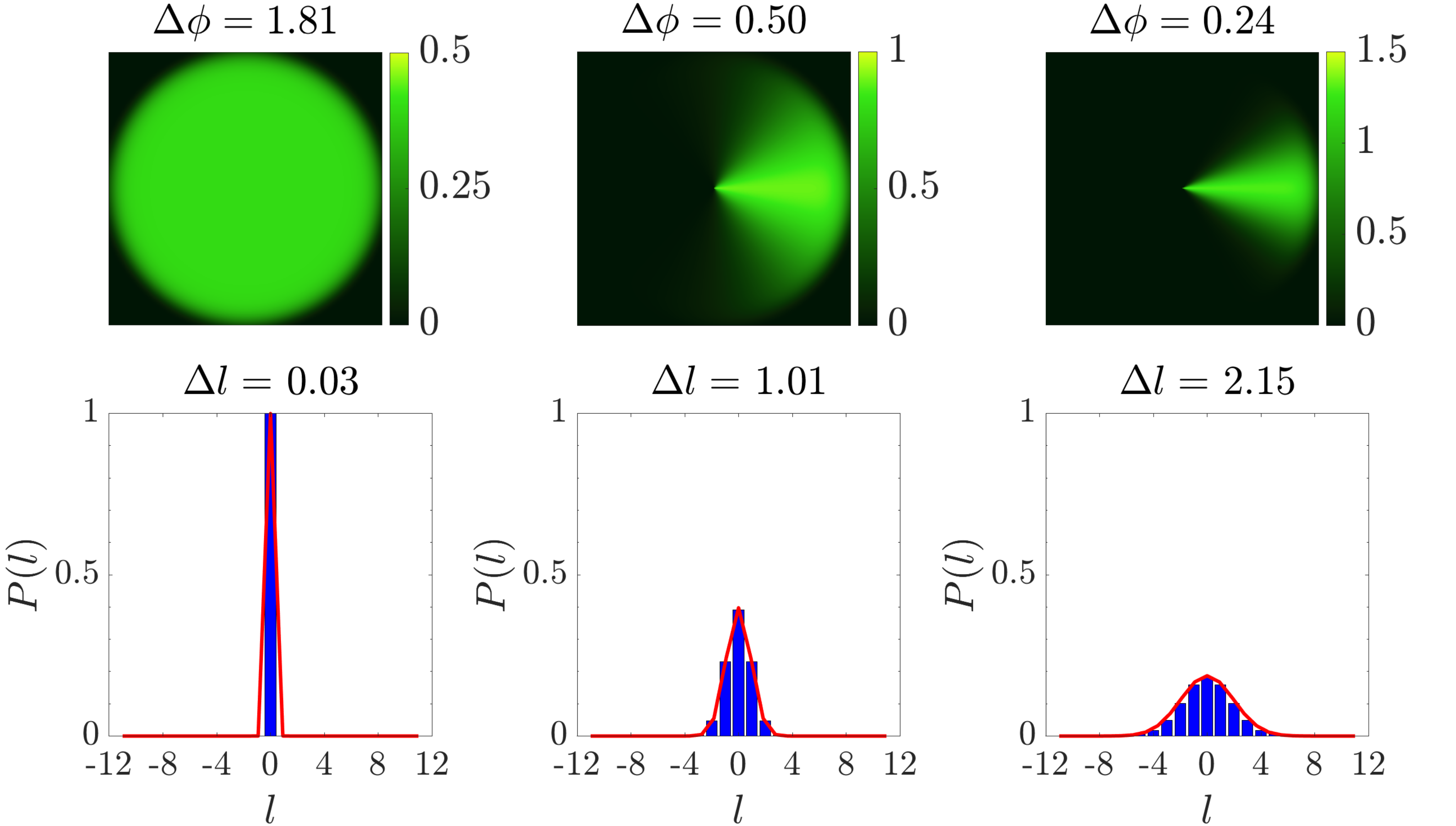}
\caption{(top row) Aperture functions $t(\rho,\phi)$ as defined in equation \eref{cake-slice} with the corresponding OAM spectra (bottom row) of the diffracted modes $\psi_0(\rho,\phi,z)$, for $\bar{l}=0$. 
The red lines in the bottom panels depict the intelligent states \eref{OAM_int_state} which best fit the OAM spectra of the shaped mode, with vanishing mean of the angular momentum, as represented by the blue histograms.
The angular and OAM uncertainties $\Delta\phi$ and $\Delta l$ (see equations \eref{angle_uncertainty} and \eref{OAM_uncertainty}), respectively, as imposed by $t(\rho,\phi)$  and inherited by diffracted state are indicated on the top of the corresponding panels.}
\label{Fig:aperture_spectra}
\end{figure*}
%%%%%%%%%%%%%%%%%%%%%%%%%%%%%%%%%%%

To ensure the normalization of the transmitted field for $z\geq 0$, we need to compensate for the losses induced by obstructing part of the beam with the aperture, and redefine
\begin{equation}
\psi_{l_0}(\rho,\phi, 0) := \frac{\psi_{l_0}(\rho,\phi, 0)}{\left[ \int\int |\psi_{l_0} (\rho,\phi, 0)|^2 \rho \rmd \rho \rmd \phi\right]^{1/2}}.
\label{renormalization}
\end{equation}
Finally, the field behind the diffracting screen is obtained by using the angular spectrum propagator 
$\psi_{l_0}(\rho,\phi, z) = \mathcal{F}^{-1} \lbrace T(\kappa, z) \mathcal{F} \left[\psi_{l_0}(\rho,\phi, 0)\right] \rbrace$, 
where ($\mathcal{F}^{-1}$) $\mathcal{F}$ is the (inverse) Fourier transform in the transverse plane, and $T(\kappa, z) = \exp [ i k z - \kappa^2 z/2k]$, with $k$ the wave number and $\kappa$ the magnitude of transverse component of the wave vector of the field, is the angular spectrum transmission function \cite{goodman}. 

\section{Uncertainty relation for angular position and angular momentum}
\label{Sec:Angle-OAM-uncertainty}
For a cylindrically symmetric problem, all physical properties are periodic functions of the azimuthal angle $\phi$. 
Accordingly, the values of a quantum mechanical observable associated with $\phi$ must lie within a $2\pi$ interval, and its canonically conjugated observable, the angular momentum operator $L_z$, can only take discrete values $\hbar l$, with $l$ an arbitrary integer number \cite{Barnett:1990}.
If we set $\phi \in [-\pi, \pi]$, the uncertainty relation for angular position and angular momentum takes the form \cite{Barnett:1990}
\begin{equation}
\Delta \phi \Delta L_z \geq  \frac{\hbar}{2}\left|1 - 2\pi P(\pi) \right|,
\label{uncertainty_relation}
\end{equation}
where $\Delta \phi$ and $\Delta L_z = \hbar \Delta l$ are, respectively, the angular position and angular momentum uncertainties, while $P(\pi)$ is the angular probability density evaluated at the boundary of the chosen $\phi$ range. 

The angular wave function of intelligent states \cite{Aragone_1974}, i.e., those states for which equality holds in equation \eref{uncertainty_relation}, can be obtained \cite{merzbacher1998} as a solution of the differential equation
\begin{equation}
\left[ i\frac{\partial }{\partial \phi} + \bar{l} + i\lambda\phi\right]g(\phi) =0,
\label{diff_eq}
\end{equation}
where $\bar{l} = \langle L_z \rangle/ \hbar$, with $\langle L_z \rangle$ the mean value of the angular momentum operator.
Equation \eref{diff_eq} is solved by the truncated Gaussian 
\begin{equation}
g(\phi) = \frac{\left(\lambda/\pi\right)^{1/4}}{\sqrt{\mathrm{erf}\left( \pi \sqrt{\lambda}\right)}} e^{i \bar{l} \phi}e^{-\lambda\phi^2/2},
\label{intelligent}
\end{equation}
which is normalized according to $\int_{-\pi}^{\pi} \rmd \phi | g(\phi)|^2 =1$.
Fourier transform of equation \eref{intelligent} yields the orbital angular momentum representation of intelligent states,
\begin{eqnarray}
g(l) &=\frac{1}{\sqrt{2\pi}}\int_{-\pi}^{\pi}e^{-i l \phi} g(\phi) \rmd \phi \nonumber \\&=\frac{(\lambda\pi)^{-1/4}}{\sqrt{\mathrm{erf}(\pi\sqrt{\lambda})}}\int_{-\infty}^{\infty}\mathrm{sinc}(k\pi) e^{-\frac{(l+k-\bar{l})^2 }{2\lambda}} \rmd k.
\label{OAM_int_state}
\end{eqnarray}

From equations \eref{intelligent} and \eref{OAM_int_state}, the uncertainties \cite{Barnett:1990},
\begin{equation}
\left( \Delta \phi \right)^2 =
%\int_{-\pi}^\pi \phi^2 |g(\phi)|^2 \rmd \phi  =
\left( \frac{1}{2\lambda} -\frac{\sqrt{\pi}e^{-\pi^2\lambda}}{\sqrt{\lambda}\mathrm{erf}(\pi \sqrt{\lambda})}\right),
\label{angle_uncertainty}
\end{equation}
and 
\begin{equation}
\left( \Delta l \right)^2 =  | \lambda |^2 \left( \Delta \phi \right)^2,
\label{OAM_uncertainty}
\end{equation}
are derived \cite{Barnett:1990},  with 
\begin{equation}
\Delta \phi \Delta l = |1 - 2\pi P(\pi)|/2,
\end{equation} 
as required for an intelligent state.

In \cite{Franke_Arnold_2004}, the uncertainty relation {\eref{uncertainty_relation} was confirmed experimentally, by observing the diffraction of the fundamental Gaussian mode on angular apertures, which match the intelligent state's wave function \eref{intelligent}.
Here, we numerically simulate (see section \ref{Sec:angular_apertures}) the diffraction of LG modes \eref{LG} on apertures described by the transmission function \eref{cake-slice}.
The angular confinement induced by the aperture induces  broadening of the OAM probability density $P(l)$ (also known as ``OAM spectrum'') of the diffracted modes, which, 
according to the uncertainty relation \eref{uncertainty_relation} is the narrower the wider the angular aperture (see figure \ref{Fig:aperture_spectra}).
\added{Similar results can be obtained by considering sharp-edge angular apertures, with the otherwise identical OAM spectra garnished by small side-lobes \cite{JhaPRA2008,Jack_NJP_2008}. }

To characterize the OAM spectrum of the diffracted states, we fitted them with the modulus square $|g(l)|^2$ of the intelligent state \eref{OAM_int_state}. 
The excellent agreement between the numerical OAM distributions and the fitting function confirms that the diffracted waves are intelligent states. 
In the following, we exploit this fact to deduce our universal law of entanglement losses induced by diffraction on angular apertures.

\section{Diffraction-induced entanglement loss}
\label{Sec:entanglement_losses}
Let us now consider the following bi-photon scenario: two single-photon excitations of LG modes with opposite OAM, $\pm \hbar l_0$, are generated in the maximally entangled Bell state
\begin{equation}
\label{psi0}
\ket{\Psi_0} = \frac{1}{\sqrt{2}}(\ket{l_0}\otimes\ket{-l_0} + \ket{-l_0}\otimes\ket{l_0}),
\end{equation}
with $\ket{\pm l_0}$ the state associated with the modes $u_{\pm l_0}$ as given in equation \eref{LG}.
The two photons then propagate through independent paths and encounter two identical, angular apertures, as defined by equation \eref{cake-slice}. 
We assume the diffracting screens to be centred with respect to the beam axis and to be placed at the beam waist $z=0$.

Diffraction on the angular apertures maps the bi-photon state \eref{psi0} onto 
\begin{eqnarray}
\Psi(\boldsymbol{\rho}_1,\boldsymbol{\rho}_2)&= \frac{1}{\sqrt{2(1+b^2)}}  \left [\psi_{l_0}(\boldsymbol{\rho}_1)\psi_{-l_0}(\boldsymbol{\rho}_2) \right.\nonumber\\ &\left. +\psi_{-l_0}(\boldsymbol{\rho}_1)\psi_{l_0}(\boldsymbol{\rho}_2) \right ],\label{diff_2photon}
\end{eqnarray}
where $\psi_{\pm l_0}(\boldsymbol{\rho})$ are the diffracted images of the encoding modes \eref{LG}, renormalized according to equation \eref{renormalization}.\footnote{\added{We omit the $z-$dependence in equation \eref{diff_2photon}, because propagation does not affect the OAM content of $\psi_{\pm l_0}$,  and, consequently, the entanglement of the biphoton state \cite{diff_paper}.}}
The extra normalization factor $\sqrt{2(1+b^2)}$, with $b$ defined as
\begin{equation}
b \equiv \int \int \psi^{*}_{-l_0}(\boldsymbol{\rho})\psi_{l_0}(\boldsymbol{\rho}) \rmd \boldsymbol{\rho},
\label{mutual_overlap}
\end{equation}
is needed because of the non-orthogonality of the diffracted fields $\psi_{-l_0}$ and $\psi_{l_0}$ (which is due to the diffraction-induced broadening of the OAM spectrum that we discussed in section \ref{Sec:Angle-OAM-uncertainty}). 
The mutual overlap $b$ ranges between $0 \leq |b| \leq 1$, where the upper bound is saturated for $\psi_{-l_0} = \psi_{l_0}$, for which equation \eref{diff_2photon} reduces to a product state.

The diffraction-induced mapping of the bi-photon state \eref{psi0} onto the state \eref{diff_2photon} is therefore local and non-unitary, hence does not preserve inner products. 
In particular, diffraction on angular apertures transforms the orthogonal LG modes $u_{\pm l_0}$ \eref{LG} in the non-orthogonal fields $\psi_{\pm l_0}$. 
This loss of orthogonality is responsible for a concomitant entanglement reduction.
The latter can be quantified by concurrence \cite{Wootters:1998,Rungta:2001}, as given by \cite{diff_paper,err_diff_paper}}
\begin{equation}
\label{C}
C = \sqrt{2\left(1 - \tr[\varrho^2]\right)} = \frac{1 - b^2}{1+b^2},
\label{concurrence}
\end{equation}
where $\varrho = \tr_1 \left[\ket{\Psi}\bra{\Psi}\right]$, with $\ket{\Psi}$ the quantum state associated with the wave function \eref{diff_2photon}, is the reduced density matrix of either one of the two photons, obtained by tracing out the other. 

%In passing to the explicit expression for $b$, we point out a formal analogy between the diffracted state \eref{diff_2photon}, given by the symmetric superposition of two non-orthogonal states, and the so called entangled coherent states \cite{Sanders:1992}:  $(\ket{\alpha}\otimes\ket{-\alpha} +\ket{-\alpha}\otimes\ket{\alpha})/\mathcal{N}_\alpha$, where $\ket{\pm \alpha}$ are coherent states and $\mathcal{N}_\alpha =\sqrt{ 2[1+\exp(-4|\alpha|^2)]}$ is a normalization constant.
%The concurrence of these states is exactly given \cite{wang2001bipartite} by equation \eref{C}, with $b$ replaced by the scalar product $\braket{\alpha}{-\alpha}=\exp(-2|\alpha|^2)$. 
%We will come back on this kinship between diffracted OAM modes and coherent states hereafter.

%%%%%%%%%%%%%%%%%%%%%%%%%%%%%%%%%%%
\begin{figure}
\centering
\includegraphics[width=0.6\textwidth]{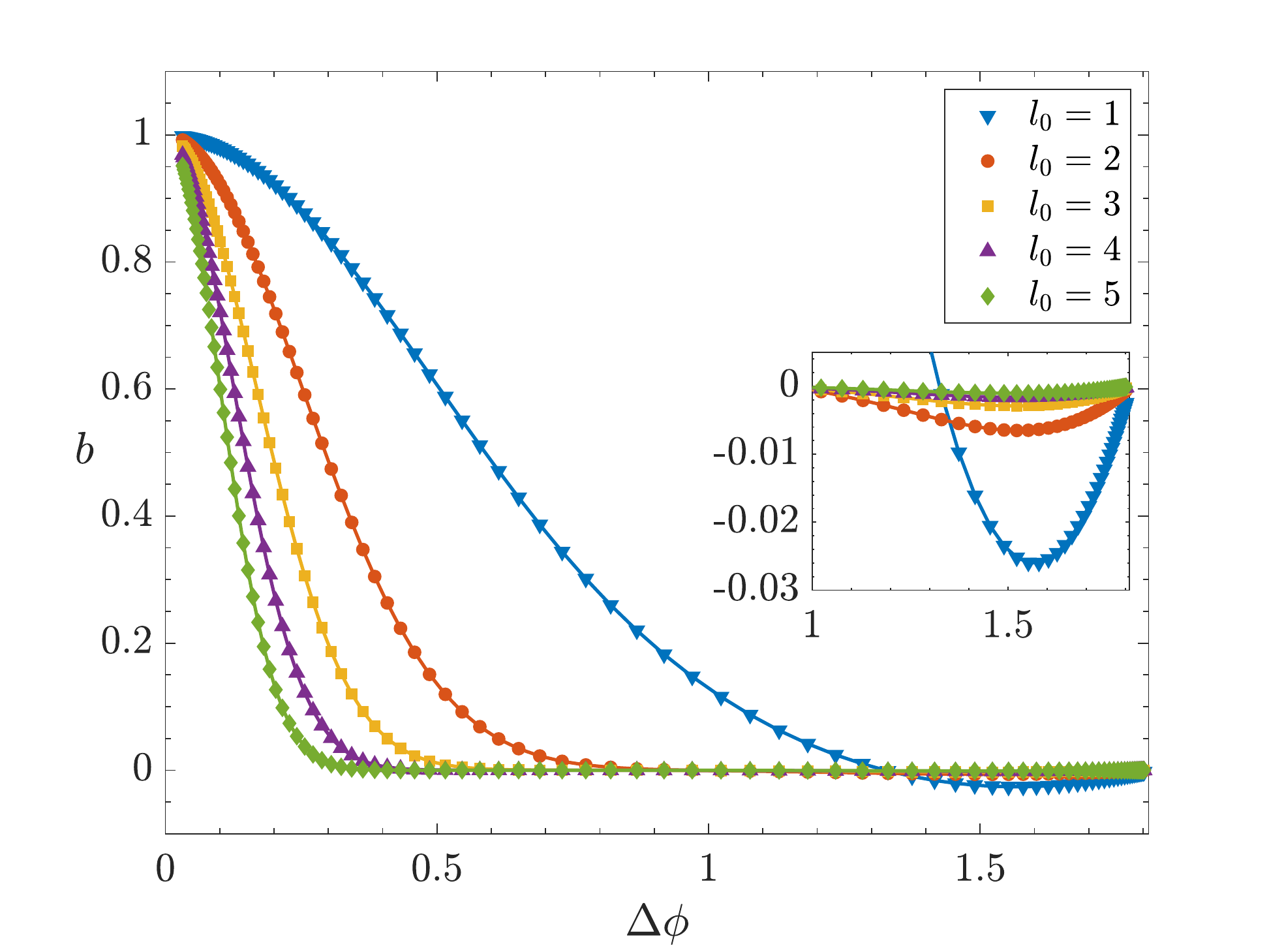}
\caption{Mutual overlap $b$ \eref{mutual_overlap} between the diffracted waves as a function of the angular uncertainty $\Delta \phi$ defined by the aperture's transmission function \eref{cake-slice} and given by equation \eref{angle_uncertainty}. 
The parameter $\lambda$ equation \eref{cake-slice} is chosen from the interval $\lambda\in [0.005,500]$.
Symbols represent numerical results, solid lines represent the analytical prediction by equation \eref{analytical_b}.}
\label{Fig:b}
\end{figure}
%%%%%%%%%%%%%%%%%%%%%%%%%%%%%%%%%%

The concurrence \eref{C} depends only on the mutual overlap $b$ of the diffracted images of the modes used to encode the entanglement in equation \eref{psi0}. 
In section \ref{Sec:Angle-OAM-uncertainty}, we have seen that these diffracted waves correspond to intelligent states. Therefore, we can use equation \eref{intelligent}, with $\bar{l}=\pm l_0$, to calculate their mutual overlap, 
\begin{eqnarray}
b &= \frac{(\lambda/\pi)^{1/2}}{\mathrm{erf}(\pi \sqrt{\lambda})}\int_{-\pi}^{\pi} e^{2 i l_0 \phi} e^{-\lambda\phi^2} \rmd \phi \nonumber\\&=  \frac{e^{-l_0^2/\lambda} \Re\left\lbrace \mathrm{erf}\left(\frac{\pi  \lambda +i l_0}{\sqrt{\lambda}}\right)\right\rbrace}{\mathrm{erf}\left(\pi  \sqrt{\lambda}\right)},
\label{analytical_b}
\end{eqnarray}
where, in the last step, we used that the real part of the error function of a complex argument is given by $\Re[\mathrm{erf}(x+iy)] = [\mathrm{erf}(x+iy) + \mathrm{erf}(x-iy)]/2$ \cite{Abrarov:2015}.
%\subsection{Universal entanglement losses at angular apertures}

The behaviour of the mutual overlap $b$ as a function of the angular uncertainty $\Delta \phi$ induced by the diffracting aperture is plotted in figure \ref{Fig:b}.
We notice that $b$ is maximal, corresponding to a bi-photon product state, in the case of diffraction on very narrow apertures with $\Delta \phi \rightarrow 0$.
As $\Delta \phi$ increases, $b$ decreases, until it reaches a minimum $b_{\mathrm{min}} <0$ (see inset in figure \ref{Fig:b}),\footnote{
We recall from equation \eref{C} that entanglement only depends on the square of $b$. Therefore, an oscillation of the mutual overlap around zero results in secondary minima of the concurrence.} and it vanishes, $b=0$, at $\Delta \phi = \pi/\sqrt{3} \simeq 1.81$, corresponding to a uniform distribution in $\phi$ \cite{JUDGE1963189}, i.e., to an open aperture. 

\begin{figure}
\centering
\includegraphics[width=0.7\textwidth]{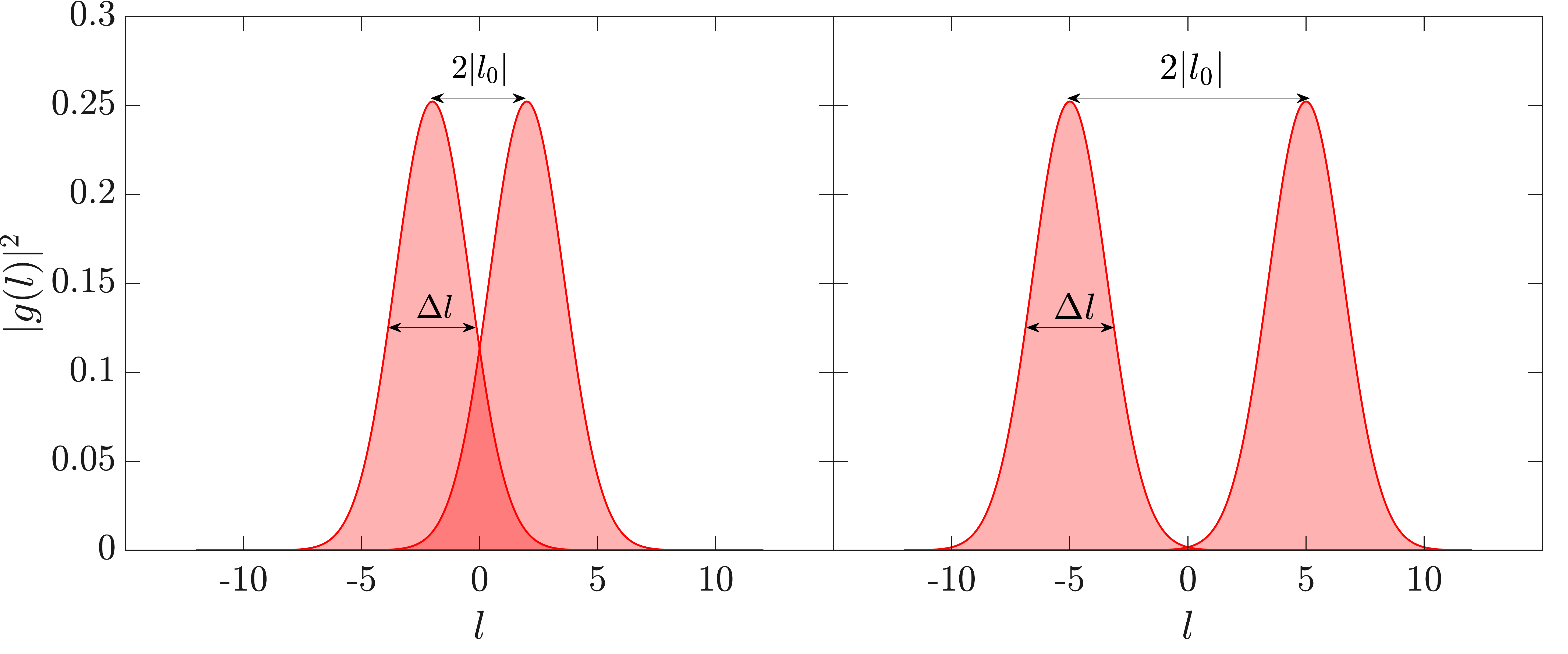}
\caption{Overlap between the wave functions of two intelligent states \eref{intelligent} with opposite average OAM ($|l_0| =2$ on the left, and $|l_0| =5$ on the right). Since $\Delta l$ is independent of $l_0$, the overlap decreases with increasing $l_0$.}
\label{Fig:overlap}
\end{figure}

We further see that the mutual overlap $b$ decays faster with $\Delta \phi$ for larger azimuthal indices $l_0$. 
To better understand this dependence on the azimuthal index, we recall that $l_0$ does not affect the uncertainty $\Delta l$, which is uniquely determined by the transmission function's \eref{cake-slice} width $\Delta \phi$.  
Consequently, the overlap $b$ between the two diffracted waves $\psi_{\pm l_0}$ decreases with the distance $2|l_0|$ between their mean values (see figure \ref{Fig:overlap}).
Accordingly, we have, from equation \eref{C}, that entanglement encoded in modes with larger OAM is more robust against angular diffraction.
%%%%%%%%%%%%%%%%%%%%%%%%%%%%%%%%%%%
%%%%%%%%%%%%%%%%%%%%%%%%%%%%%%%%%%%

\begin{figure}
\centering
\includegraphics[width=0.5\columnwidth]{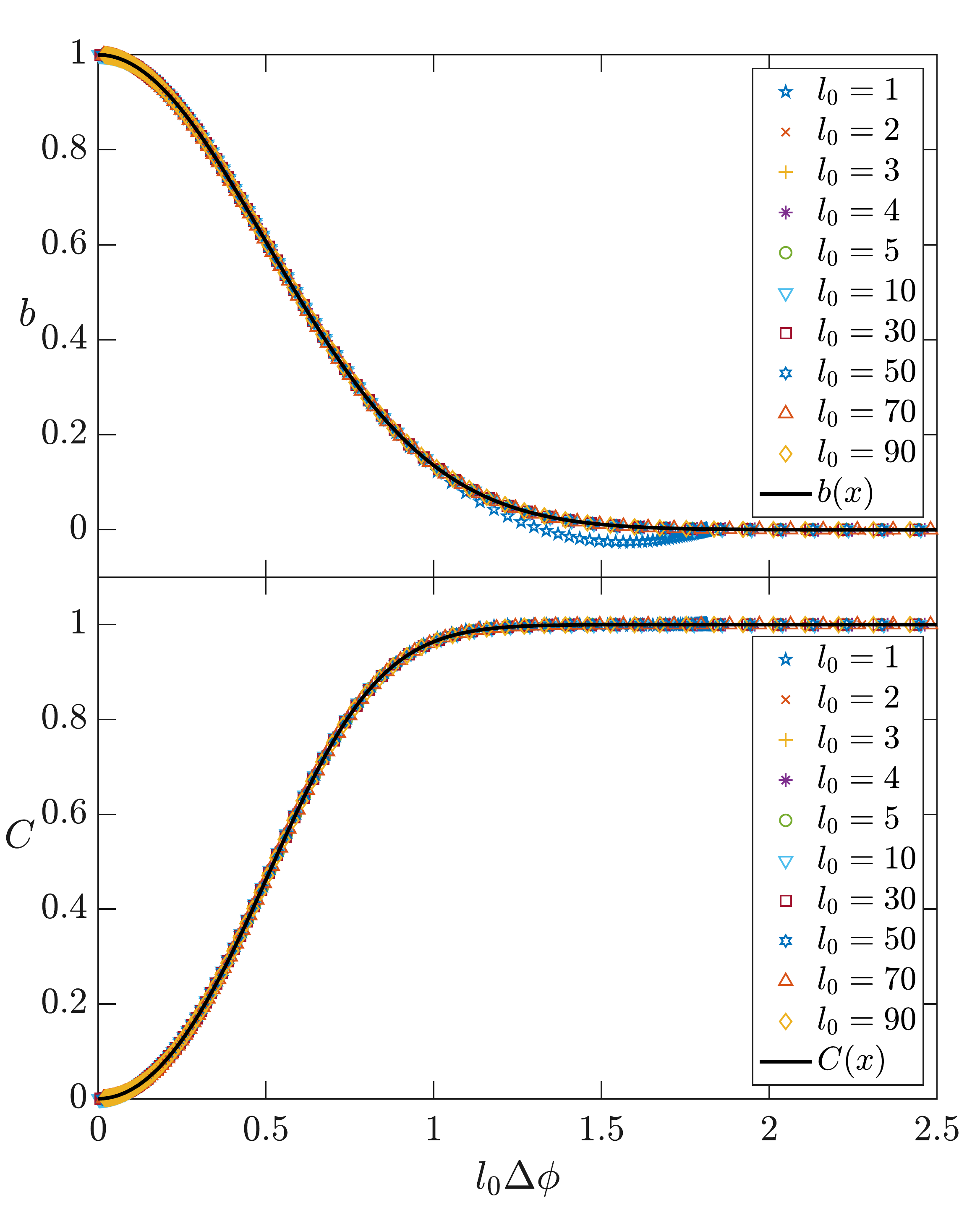}
\caption{ 
The Gaussian approximation \eref{b_universal} for the universal behaviour of the mutual overlap, and the concomitant result \eref{univ_concurrence} for concurrence (black lines in the top and bottom panel) are compared to the numerical results for different values of the input OAM value $l_0$ (coloured symbols as listed in the legends). The agreement is excellent, apart from small deviations for $l_0=1$ in the top panel.}
\label{Fig:rescaled}
\end{figure}

Equivalently, we can attribute the enhanced stability of entanglement encoded in higher-order OAM modes to their finer phase structure, determined by the angle $\pi/2|l_0|$ that also underlies the higher resilience of such modes in turbulence \cite{Leonhard2015}.
As a consequence, an angular uncertainty $\Delta \phi$ is effectively $l$ times smaller for a mode with $l_0  = l$ than for a mode with $l_0=1$.
Therefore, we expect the mutual overlap of the modes $\psi_{\pm l_0}$, and consequently the concurrence of the output state, to fall on a universal curve when we rescale the angular uncertainty $\Delta \phi$ with the azimuthal index $l_0$. 
Figure \ref{Fig:rescaled} shows that this conjecture is indeed correct.

We can obtain an analytical expression for this universal behaviour by noting that for $\lambda \gtrsim 0.5$ the exponential term in equation \eref{angle_uncertainty} is negligible. This translates to the interval $0 \leq \Delta \phi \lesssim 1$, where (for $|l_0| >1$) the mutual overlap $b$ decreases from $1$ to $0$, and we can approximate equation \eref{angle_uncertainty} by
\begin{equation}
(\Delta \phi)^2 \approx \frac{1}{2\lambda}.
\label{approx_deltaphi}
\end{equation}
Furthermore, the Gaussian factor in equation \eref{analytical_b} is significantly different from zero only for $\lambda \gtrsim l_0^2$, under which condition the error functions are in essence equal to  unity. 
Therefore, we can write
\begin{equation}
b \approx e^{-l_0^2/\lambda} \approx e^{-2(l_0\Delta\phi)^2},
\label{b_universal}
\end{equation} 
 which is the Gaussian approximation for the mutual overlap.
 
Equation \eref{b_universal} is quite noteworthy. First of all, this approximate formula yields excellent agreement with the exact solution \eref{analytical_b} for all azimuthal indices except $l_0=1$.\footnote{In the latter case, the Gaussian \eref{b_universal} fails to reproduce the negative minimum at $\Delta \phi \approx 1.55$. 
This leads to an error of $\approx 3\%$, which is not unexpected, since our approximation is valid in the range  $0 \leq \Delta \phi \lesssim 1$.} 
Second, since the right hand side of \eref{b_universal} depends on a single parameter $l_0\Delta\phi$, this equation establishes the explicit form of the universal behaviour of the mutual overlap and, hence, of entanglement. Indeed, substitution of  equation \eref{b_universal} into equation \eref{C} yields an expression which implies the universal behaviour of concurrence,
\begin{equation}
C (\ket{\Psi}) \approx \tanh \left(2 l_0^2 \Delta \phi^2 \right),
\label{univ_concurrence}
\end{equation}
which is graphically indistinguishable from the exact results (see the bottom panel in figure \ref{Fig:rescaled}).

Furthermore, this result allows us to identify a close link between the diffracted state \eref{diff_2photon}, given by the symmetric superposition of two non-orthogonal states, and superpositions of coherent states with opposite phase:  $(\ket{\alpha}\otimes\ket{-\alpha} +\ket{-\alpha}\otimes\ket{\alpha})/\mathcal{N}_\alpha$, where $\mathcal{N}_\alpha =\sqrt{ 2[1+\exp(-4|\alpha|^2)]}$ is a normalization constant.
The concurrence of these states exactly coincides \cite{wang2001bipartite} with equation \eref{univ_concurrence}, when $l_0\Delta \phi$ is replaced by $|\alpha|$. 
In this sense, the diffracted states $\psi_{\pm l_0}(\boldsymbol{\rho})$ can be approximately identified with coherent states $\ket{\alpha = l_0\Delta\phi}$.
%In other words, the diffracted states $\psi_{\pm l_0}(\boldsymbol{\rho})$ correspond to the coherent states $|\pm l_0\Delta \phi\rangle$.
 %%%%%%%%%%%%%%%%%%%%%%%%%%%%%%%%%%%
%%%%%%%%%%%%%%%%%%%%%%%%%%%%%%%%%%%

Finally, let us recall that, also in weak atmospheric turbulence, OAM qubits exhibit universal entanglement decay \cite{Leonhard2015}.
In the most optimistic (for quantum communication) scenario, the coupling between the encoding modes is an exponentially decaying function of the azimuthal index $l_0$ divided by the characteristic number of surface elements with uncorrelated phase errors in the cross-section of the propagated beam's wave front \cite{Bachmann:2019}.
Thus, the mutual overlap among the OAM modes induced by diffraction that we studied in this work decays much faster with increasing $l_0$ than the one due to atmospheric turbulence (Gaussian vs exponential decay).

\section{Conclusion}
\label{Sec:conclusion}
We studied the diffraction of LG-encoded entangled bi-photons from angular apertures.
We showed that the diffracted images of the LG modes correspond to intelligent states \eref{intelligent} which saturate the uncertainty relation \eref{uncertainty_relation} for angular position and angular momentum.
Exploiting this fact, we established a connection between diffraction-induced entanglement losses and the angular uncertainty introduced by the aperture.
We showed that the entanglement of the diffracted bi-photon state, as quantified by concurrence, is a universal function \eref{univ_concurrence} of the product of the angular uncertainty $\Delta \phi$ induced by the diffracting screen and the azimuthal index $l_0$ of the modes used to encode the entanglement. Furthermore, we demonstrated that the diffracted waves de facto are analogous to the coherent states.

These findings imply that interpreting diffraction phenomena in terms of angular confinement is useful to derive approximations and bounds on the diffraction-induced entanglement losses, by exploiting the angle-OAM uncertainty relation \eref{uncertainty_relation}.
In the future, it will be interesting to explore this idea by considering diffracting screens of arbitrary shape. %Another topic of future research is suggested by the close analogy between the spatially entangled OAM modes that are diffracted on angular apertures and the entangled coherent states. The latter decohere very rapidly, when coupled to a bath \cite{WILSON_2002}. Shall we observe a similar sensitivity of the diffracted modes to perturbations induced, for instance, by a turbulent medium?

The results of this work were derived using the analytical expression \eref{C} for the concurrence of a diffracted OAM-qubit pair, that allows to quantify the entanglement of an infinite-dimensional pure state.
A generalization of this formula to higher-dimensional input states will likely depend on multiple mutual overlaps between the different spatial modes used for the encoding.
Exploring how the interplay between these different mutual overlaps determines the entanglement losses of high-dimensional input states may shed some light on the general mechanisms that deteriorate entanglement encoded in spatial modes of light. 

\ack
We would like to thank S. M. Barnett for illuminating and enjoyable discussions. 
We acknowledge support by the Deutsche Forschungsgemeinschaft under Grant No. DFG BU 1337/17-1.  

\section*{References}
%% bibliography
\bibliography{biblio}
\bibliographystyle{iopart-num}
\end{document}